\documentstyle[11pt,newpasp,twoside,epsf]{article}
\reversemarginpar

\def\om{\Omega}     
\def\kin{{\cal V}}     
\def\pin{{\cal X}}     
\def\vpd{{\cal R}}     
\def\len{a_B}     
\def\lag{D_L}

\begin{document}
\title{Fast bars in SB0 galaxies}

\author{Enrico Maria Corsini} 
\affil{Dipartimento di Astronomia, Universit\`a di Padova,
Padova, Italy}

\author{J. A. L. Aguerri}
\affil{Instituto de Astrof\'{\i}sica de Canarias, La Laguna, Spain}

\author{Victor P. Debattista}
\affil{Institut f\"ur Astronomie, ETH H\"onggerberg, Z\"urich, Switzerland}

\begin{abstract}
We measured the bar pattern speed in a sample of 7 SB0 galaxies using
the Tremaine-Weinberg method. This represents the largest sample of
galaxies for which bar pattern speed has been measured this way. All
the observed bars are as rapidly rotating as they can be.
We compared this result with recent high-resolution $N$-body
simulations of bars in cosmologically-motivated dark matter halos, and
conclude that these bars are not located inside centrally concentrated
halos.

\end{abstract}

\noindent
The pattern speed of a bar, $\om$, is its main kinematic observable.
When pa\-ra\-me\-tri\-zed by the distance-independent ratio $\vpd
\equiv \lag/\len$ between the corotation radius, $\lag$, and the bar
semi-major axis, $\len$, it permits the classification of bars into fast
($1.0 \leq \vpd \leq 1.4$) and slow ($\vpd > 1.4$) ones. 
A model-independent method for measuring $\om$ directly was obtained
by Tremaine \& Weinberg (1984, hereafter TW).  The TW method is given
by the simple expression $\pin\,\om\,\sin i = \kin$, where $\pin$ and
$\kin$ are luminosity-weighted mean position and velocity measured
along slits parallel to the line-of-nodes.  If a number of slits at
different offsets from the major-axis are obtained for a galaxy, then
plotting $\kin$ versus $\pin$ for the different slits produces a
straight line with slope $\om \sin i$.

\begin{figure}
\plotfiddle{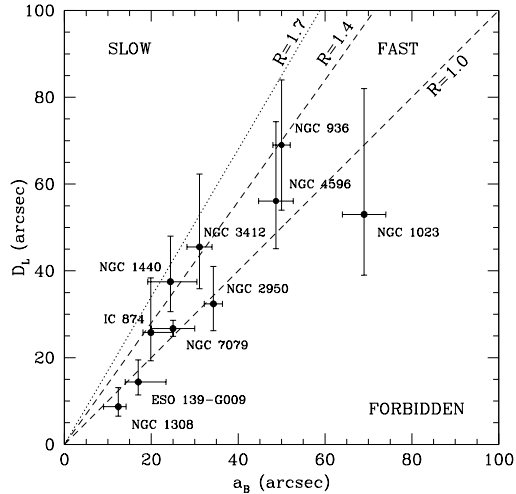}{6cm}{0}{34}{34}{-100}{-57}
\caption{The corotation radius, $D_{L}$, and the bar semi-major axis, 
  $a_{B}$, for the sample galaxies, NGC 936 (Merrifield \& Kuijken
  1995), NGC 4596 (Gerssen et al. 1997) and NGC 7079 (Debattista \&
  Williams 2003). Dashed lines corresponding to $\vpd=1$ and
  $\vpd=1.4$, separate the fast-bar, slow-bar and forbidden
  regimes. }
\end{figure}

We defined a sample of 7 SB0 galaxies with intermediate inclination, a
bar at intermediate position between the major and minor axes of the
disk and no evidence of spirals and patchy dust (NGC 1023: Debattista
et al. 2002; ESO 139-G9, IC 874, NGC 1308, NGC 1440, and NGC 3412:
Aguerri et al. 2003; NGC 2950: Corsini et al. 2003). For each sample
galaxy we obtained $I-$band imaging and long-slit spectra with slits
parallel to disk major axis.
We measured $\om$ with the TW method. The corotation radius, $\lag
\equiv V_{\rm c}/\om$, was derived from the circular velocity, 
$V_{\rm c}$, after applying the asymmetric drift correction to the
stellar velocities and velocity dispersions.
The len\-gth of the bar, $a_B$, was de\-rived from the analysis of the
surface brightness distribution.
For all of the sample galaxy $\vpd$ is consistent with being in the
range 1.0 to 1.4, within the errors, i.e.  with each having a fast bar
(Fig. 1).  The unweighted average for the sample is $\overline{\vpd} =
1.1$. Some of the values of $\vpd$ are nominally less
than unity, this suggests that the large range of $\vpd$ is a
result of random errors and/or scatter.

Debattista \& Sellwood (2000) argued that bars this fast can only
survive if the disc in which they formed is maximal.  Recent high
resolution $N$-body simulations with cosmologically-motivated dark
matter halos produce bars with $\vpd$ as large as 1.7 (Valenzuela \&
Klypin 2003). Even discounting our argument above in favor of a more
restricted range of $\vpd$, Fig.  1 shows that $\vpd = 1.7$ is
possible only for the bars of IC 874, NGC 1440, NGC 3412 and,
marginally, NGC 936 (Merrifield \& Kuijken 1995), while the bars of
ESO 139-G009, NGC 1023, NGC 1308, NGC 2950, NGC 7079 (Debattista \&
Williams 2003) and NGC 4596 (Gerssen et al. 1997) never reach this
value of $\vpd$. Therefore we conclude that the $N$-body models of
Valenzuela \& Klypin (2003) probably produce slower bars than the
observed.

\end{document}